\documentstyle[preprint,aps]{revtex}

\tightenlines
\begin{document}

\draft
\preprint{}
\title{A New Prediction for Direct CP Violation $\varepsilon'/\varepsilon$ \\ and
$\Delta I = 1/2$ Rule  }
\author{ Yue-Liang  Wu  }
\address{Institute of Theoretical Physics, Academia Sinica, \\
 P.O. Box 2735, Beijing 100080, P.R. China }
\maketitle

\begin{abstract}
The low energy dynamics of QCD is investigated with special
attention paid to the matching between QCD and chiral perturbation
theory(ChPT), and also to some useful algebraic chiral operator
relations which survive even when we include chiral loop
corrections. It then allows us to evaluate the hadronic matrix
elements below the energy scale $\Lambda_{\chi} \simeq 1$ GeV.
Based on the new analyzes, we present a consistent prediction for
both direct CP-violating parameter $\varepsilon'/\varepsilon$ and
$\Delta I =1/2$ rule in the kaon decays. In the leading $1/N_c$
approximation, the isospin amplitudes $A_0$ and $A_2$ are found to
agree well with the data, and the direct CP-violating parameter
$\varepsilon'/\varepsilon$ is predicted to be large, which also
confirms our early conclusion. Its numerical value is
$\varepsilon'/\varepsilon = 23.6^{+12.4}_{-7.8}\times
10^{-4}\left( Im\lambda_t/1.2\times 10^{-4}\right)$ which is no
longer sensitive to the strange quark mass due to the matching
conditions. Taking into account a simultaneous consistent analysis
on the isospin amplitudes $A_0$ and $A_2$, the ratio
$\varepsilon'/\varepsilon$ is in favor of the values
$\varepsilon'/\varepsilon = (20\pm9)\times 10^{-4}$.
\end{abstract}

\newpage

\section{Introduction}

  To make a consistent prediction for the direct CP-violating parameter
$\varepsilon'/\varepsilon$ caused by the Kobayashi-Maskawa
CP-violating phase\cite{KM} which can arise from the explicit CP
violation in the standard model(SM) or originate from the
spontaneous CP violation\cite{TDL} in the simple extension of SM
with two Higgs doublets (S2HDM) \cite{WW}, it is thought to be
necessary to understand simultaneously the longtime puzzle of the
$\Delta I = 1/2$ rule in kaon decays as they involve the
long-distance evolution of common hadronic matrix elements. It is
believed that the low energy dynamics of QCD shall play a crucial
role for a consistent analysis. During the past few years,
 both theoretical and experimental efforts on direct CP violation in the kaon
 decays have been made important progresses. As a consequence,
 it has reached an agreement between the experimental
 results\cite{NA48N,NA31,E731,KTEV,NA48} and the theoretical
 predictions within the framework of ChPT\cite{YLW1,HPSW}
 and chiral quark model\cite{BEFL}. On the experimental side, two improved new
 experiments\cite{KTEV,NA48} with higher precision have reported results which are
 consistent each other and also agree with the early result\cite{NA31}. On the theoretical
 side, there have been some developments which are mainly based on QCD of quarks and
 cut-off ChPT at low energies for mesons.
 The renormalization coefficients of all the relevant four quark operators,
 which characterize the short-distance effects of the effective Hamiltonian generated
 from renormalization of the weak interactions, has been computed and extended
from the leading order\cite{BBH,PW} to the next-to-leading
 order\cite{BJLW,CFMR} QCD corrections. The results agree with each other. The long-distance
 effects have been evaluated from the ChPT inspired from $1/N_c$ expansion \cite{THOOFT,EW}
 up to the chiral one-loop level\cite{BBG,YLW1,HPSW} as well as
 from chiral quark model\cite{BEFL}. Recently they have been recalculated within the
 same framework of ChPT but with a different calculating scheme\cite{FG,DM1,DM2,DM3}.
 The important issue concerned in all the calculations
 is the matching problem to QCD. In ChPT, it requires the matching between the
 short-distance operator evolution from QCD with infrared cut-off and long-distance
 operator evolution from ChPT with ultraviolet cut-off.
 Practically, the renormalization scale $\mu$ dependence of the
 Wilson coefficient functions $c_{i}(\mu)$ from perturbative QCD
 should cancel the one of the corresponding operators $Q_i(\mu)$
 from non-perturbative contributions. In the chiral quark model the
 operator evolution leads to the results which are expected to
 valid only at a special value of the energy scale $\mu$.
 Alternatively, in the ChPT approach, its attractive advantage is
 that chiral loops with an ultraviolet cut-off, denoted by $M$,
 introduce a scale-dependence for
 long-distance operator evolution. As a simple consideration, the ultraviolet cut-off $M$
 might naively be identified to the infrared cut-off $\mu$ to improve the matching.
 Consequently, both the $\Delta I = 1/2$ rule and direct CP-violating parameter
 $\varepsilon'/\varepsilon$ can be enhanced to be more consistent with the present experimental
 data. Nevertheless, in the existing treatments of the approach, there remain some open
 questions which need to be further clarified. Firstly, the momentum cut-off $M$ in
 the long-distance operator evolution from meson loops can in general only be extended to
 the energy scale which must be smaller than the dynamical chiral symmetry breaking scale
 $\Lambda_{f}$, i.e. $M < \Lambda_{f}\sim$ 1GeV, whereas the short-distance
 operator evolution from perturbative QCD (by using renormalization group equation)
 requires that the energy scale should be above the confining scale, i.e., $\mu >$ 1 GeV.
 Thus naively identifying the ultraviolet cut-off $M$ in ChPT to the infrared cut-off $\mu$
 in perturbative QCD may become inappropriate. Secondly, there appear some
 discrepancies between the ref.\cite{BBG} and ref.\cite{FG,DM2}
 for the matrix elements $Q_1$ and $Q_2$ even if the same chiral Lagrangian has been
 used and the same loop diagrams have been considered. It is noticed that the
 discrepancies only occur in the coefficients of the quadratic terms of the cut-off
 energy scale $M$ and in the constant terms. Such discrepancies mainly arise from
 different calculating schemes. In refs.\cite{BBG,YLW1,HPSW}, all the chiral one-loop
 contributions were considered to be summed up with the cut-off regularization, and
 the coupling constants are replaced by the renormalized ones. Such a
 treatment is the standard one as adopted in the quantum field theory. In the recent
 calculations\cite{DM1,DM2,DM3}, the chiral one-loop diagrams have been separated into
 two classes, i.e., so-called factorized and nonfactorized diagrams.
 For the nonfactorized diagrams, a virtual momentum flow has been artificially
 added to the propagators. As a consequence, such an alternative treatment is equivalent to the
 change of cut-off energy scale. When taking the cut-off energy scale to be
 infinity as the case for a renormalized field theory like QCD and QED,
 the treatment has no effects. However, for a finite cut-off integral, the change
 of the variables of the integrand will result in different results.
 This is the main reason why two calculations led to different
 results in the quadratic terms of cut-off scale and in the constant terms. In fact,
 one can simply rescale the cut-off scale $\Lambda_c$ in refs.\cite{DM2} via
 $\Lambda_c^2 = 2M^2 /3$ to obtain the results in ref.\cite{BBG} for $Q_1$ and $Q_2$.
 With a similar reason, for $Q_6$ there also exist discrepancies between ref. \cite{YLW1,HPSW}
 and ref. \cite{DM3} for the quadratic and constant terms, however, two results cannot be
 simply related by the same rescaling factor as the one for $Q_1$ and $Q_2$.
 The reason is that the results in ref.\cite{DM3} were obtained only by evaluating
 part of the so-called nonfactorized diagrams. Notice that it may not be
 so clear to separate the factorized and nonfactorized diagrams for the loop corrections of
 the left-right structure four quark operators generated from the penguin diagrams in which
 the intermediate quarks form a closed loop with the gauge bosons.
 In fact, for the density $\times$ density operators, the so-called
 factorized diagrams do provide contributions to the anomalous dimension of the operators
 in QCD evolution. The $\mu$-dependence of such factorized contributions is exactly
 cancelled by the one of explicit quark mass factor appearing in the corresponding
chiral operators,  but not by
 the one of the corresponding factorized chiral loop, since the quark mass factor does not arise
 from the chiral loop contributions. Therefore, for the density $\times$ density operators,
 or more general for the operators with left-right structure, it is not necessary to have
 one to one correspondings between QCD loop and chiral loop due to
 the $\mu$-dependence of the low energy coupling constants
 in the chiral Lagrangian. Nevertheless, physics observables should be independent of
 the calculating schemes.

 Thus, clarifying the above two open questions comes to one of the
 main purposes of this paper. Our paper is organized as follows: in section 2, we describe
 the basic motivations for evaluating the long-distance contributions of the hadronic
 matrix elements within the framework of ChPT. Especially, a functional cut-off momentum
is introduced for the purpose of matching, namely the cut-off momentum $M$ is in general
 considered as the function of the QCD running scale $\mu$, i.e., $M\equiv M(\mu)$;
 in section 3 we explicitly write down the chiral representation of four qurak operators
 and emphasize some useful algebraic chiral operator relations;
 in section 4 we investigate the matching between QCD and ChPT, where the
 chiral operators are explicitly evaluated in the functional cut-off momentum scheme,
 two useful matching conditions will be obtained. Of interest, the strange quark mass is
 found to be fixed from the matching condition and algebraic chiral operator relation.
 The long-distance chiral operator evolution is carried out in section 5; in section 6
 we present our numerical predictions for the direct CP-violating parameter
 $\varepsilon'/\varepsilon$ and $\Delta I = 1/2$ rule, they are found to be remarkably
 consistent with the data. Our conclusions and remarks are made in the last section.

\section{Basic Motivation}

 Our considerations are mainly based on the following basic points:

\begin{itemize}
\item In the large $N_c$ limit but with the combination $\alpha_s N_c \equiv \alpha_0$
       being held fixed. The QCD loop corrections which are proportional to
       $\alpha_s$ are then corresponding to a large $N_c$ expansion, $\alpha_s \sim 1/N_c$
       \cite{THOOFT}.
\item Chiral symmetry is supposed to be broken dynamically due to attractive gauge
       interactions, namely the chiral condensates $<\bar{q}q> $
        exist and lead to the Goldstone-like pseudoscalar mesons $\pi$, $K$,
       $\eta$. The chiral symmetry breaking scale $\Lambda_{f}$ is characterized by the condensate,
       $\Lambda_{f} \simeq 4\pi \sqrt{-2<\bar{q} q> /r } \sim 1$ GeV
        with $r = m_{\pi_0}^{2}/\hat{m}$ ($\hat{m} = (m_u + m_d)/2 $).
\item The chiral Lagrangian is considered to describe the low energy dynamics of
       QCD in large $N_c$ limit and is going to be treated as a cut-off effective field theory.
       The cut-off momentum $M$ is expected to be below the chiral symmetry breaking scale
       $\Lambda_{f}$.
\item The chiral meson loop contributions are characterized by the powers of
       $p^{2}/\Lambda_{f}^{2}$ with $\Lambda_{f}= 4\pi f$ .
       Here $f^{2} \simeq -2<\bar{q} q> /r \sim N_c$ is at the leading $N_c$ order and fixed by
       the $\pi$ decay coupling constant $f\sim F_{\pi}$.
       Thus the chiral meson loop contributions are also corresponding to a
       large $N_c$ expansion of QCD, $p^{2}/\Lambda_{f}^{2} \sim 1/N_c\sim \alpha_s$.  Therefore
       both chiral loop and QCD loop contributions must be matched each other, at least in
       the sense of large $N_c$ limit. Thus the final physical results should be
       independent of the cut-off schemes.
 \item The cut-off momentum $M$ of loop integrals should not be naively identified to
       the renormalization scale $\mu$ appearing in the perturbative QCD in large $N_c$ limit.
       It is in general taken to be a function of $\mu$, i.e., $M \equiv M(\mu)$,
       which may be regarded as a functional cut-off momentum, its form is determined by
       the matching between the Wilson coefficients of QCD and hadronic matrix elements
       evaluated via ChPT.  It is seen that  the matching relates the chiral cut-off momentum
       to the strong coupling constant so that the results become scheme independent.
  \end{itemize}

  From these points of view, the ChPT with functional cut-off momentum
 is going to be treated, in certain sense, as a low energy effective field
 theory of QCD in the large $N_c$ limit. With such a treatment, it is in
 general not necessary to distinguish the so-called factorized and
non-factorized contributions since the renormalization of field
theory should well cover both of their contributions
automatically. In this paper, we will give up the calculating
scheme of separating the factorized and nonfactorized
contributions, and adopt the calculating scheme first proposed by
Bardeen, Buras and G\'{e}rard\cite{BBG} in the ChPT inspired by
the $1/N_c$ expansion, but with a functional cut-off momentum
$M(\mu)$ instead of naively
 identifying the cut-off momentum to the QCD running scale $\mu$.

\section{Chiral Representation and Algebraic Relations}

  In the standard model, the $\Delta S = 1$ low energy ($\mu < m_c$) effective Hamiltonian
  for calculating $K\rightarrow \pi \pi$ decay amplitudes can be written as

\begin{equation}
{\cal H}_{eff}^{ \Delta S=1}= \frac{G_F}{\sqrt{2}}
\;\lambda_u\sum_{i=1}^8 c_i(\mu)\,Q_i(\mu)\  ,
 \quad ( \mu < m_c)
\end{equation}
with $Q_i$ the four quark operators
\begin{eqnarray}
Q_1 &=& 4\,\bar{s}_L\gamma^{\mu} d_L\,\bar{u}_L\gamma_{\mu} u_L\, , \
\hspace{2.2cm}
Q_2 \,\,\,=\,\,\, 4\,\bar{s}_L\gamma^{\mu} u_L\,\bar{u}_L
\gamma_{\mu} d_L\,, \label{qia} \nonumber \\
Q_3 &=& 4\,\sum_q \bar{s}_L\gamma^{\mu} d_L\,\bar{q}_L\gamma_{\mu} q_L\,,
\hspace{1.75cm}
Q_4 \,\,\,=\,\,\, 4\,\sum_q \bar{s}_L\gamma^{\mu} q_L\,\bar{q}_L
\gamma_{\mu} d_L\,, \nonumber \\
Q_5 &=& 4\,\sum_q \bar{s}_L\gamma^{\mu} d_L\,\bar{q}_R\gamma_{\mu} q_R\,,
\hspace{1.75cm}
Q_6 \,\,\,=\,\,\, -8\,\sum_q \bar{s}_L q_R\,\,\bar{q}_R d_L\,,\\
Q_7 &=& 4\,\sum_q \frac{3}{2}e_q\,\bar{s}_L\gamma^{\mu} d_L\,
\bar{q}_R \gamma_{\mu} q_R\,,\hspace{1.0cm}
Q_8 \,\,\,=\,\,\, -8\,\sum_q \frac{3}{2}e_q\,\bar{s}_L q_R\,
\bar{q}_R d_L\,, \nonumber
\end{eqnarray}
with $q_{R,L}=\frac{1}{2}(1\pm\gamma_5)q $. Where the sum goes
over the light flavors ($q=u,d,s$) and $e_q$ is the charge of the
corresponding light quarks. $Q_3,\ldots,Q_6$ arise from strong
penguin diagrams. They transform as $(8_L,1_R)$ under
$SU(3)_L\times SU(3)_R$ and solely contribute to $\Delta I=1/2$
transitions. Note that only seven operators are independent as the
linear relation $Q_4 = Q_2 - Q_1 + Q_3$.  $Q_7$ and $Q_8$
originate from electroweak penguin diagrams. $c_i(\mu)$ are Wilson
coefficient functions
\begin{equation}
c_i(\mu)=z_i(\mu)+\tau y_i(\mu)\ .
\end{equation}
where $\tau=-\lambda_t/\lambda_u $ with $\lambda_q=V_{qs}^*\,V_{qd}$. The Wilson
coefficient functions  $z_i(\mu)$ and $y_i(\mu)$ have been evaluated up to
the next-to-leading order QCD corrections.
The $K\rightarrow \pi \pi$ decay amplitudes $A_I$ with isospin $I$ are given by
 \begin{equation}
 A_I e^{i\delta_{I}} = \langle \pi\pi | {\cal H}_{ef\ f}^{ \Delta S=1} | K \rangle \equiv
 \frac{G_F}{\sqrt{2}} \;\lambda_u\sum_{i=1}^8 c_i(\mu)\langle Q_i(\mu)\rangle_I
 \end{equation}
 where $\delta_{I}$ are the final state strong interaction phases.
It is a hard task for calculating the hadronic matrix elements $
\langle Q_i(\mu) \rangle_I $ for $\mu < \Lambda_{\chi}= 1$ GeV
which is at the order of chiral symmetry breaking scale. This is
because perturbative QCD becomes unreliable in such a low energy
scale. In this paper we adopt the ChPT with functional cut-off
momentum to evaluate $ \langle Q_i(\mu) \rangle_I $ when $\mu <
\Lambda_{\chi}$. To do that, the procedure is as follows: one
first represents the current $\times$ current or density $\times$
density four quark operators $Q_i$ by bosonized chiral fields from
the chiral Lagarangian, then calculate loop contributions by using
the functional cut-off momentum scheme. Finally, one matchs the
two results obtained from QCD and ChPT with functional cut-off
momentum
 by requiring scale independence of the physical results.

 The general form of the chiral Lagrangian can be expressed
 in terms of the expansions of the momentum $p$ and quark mass
 to the energy scale $\Lambda_{\chi}$. Here we only use the chiral
 Lagrangian which is relevant to the $K\rightarrow \pi\pi$
 decays (for the most general one, see ref.\cite{CHPT})
  \begin{eqnarray}
{\cal L}_{eff}&=&\frac{f^2}{4}\{ \  tr(D_{\mu} U^{\dagger}
D^{\mu} U ) + \frac{m_{\alpha}^2}{4N_c} tr( \ln U^{\dagger} -\ln U
)^2 +r\ tr({\cal M} U^{\dagger}+U{\cal M}^{\dagger} ) \nonumber \\
& & + r \frac{\chi_5}{\Lambda_{\chi}^2} tr\left( D_{\mu}
U^{\dagger} D^{\mu}  U({\cal M}^{\dagger} U + U^{\dagger}{\cal M}
\right) \\
&& + r^2 \frac{\chi_8}{\Lambda_{\chi}^2} tr\left( {\cal
M}^{\dagger} U{\cal M}^{\dagger} U+{\cal M} U^{\dagger}{\cal M}
U^{\dagger} \right) + r^2 \frac{\kappa_2}{\Lambda_{\chi}^2} tr
({\cal M}^{\dagger} {\cal M} )\  \} \nonumber
\end{eqnarray}
with $D_{\mu} U=\partial_{\mu} U-ir_{\mu} U + iU l_{\mu}$, and
${\cal M}=\mbox{diag}( m_u,m_d,m_s)$. $l_{\mu}$ and $r_{\mu}$ are
left- and right-handed gauge fields, respectively. The unitary
matrix $U$ is a non-linear representation of the pseudoscalar
meson nonet given as $U= e^{i\Pi /f}$ with $\Pi=\pi^a\lambda_a$
and $tr(\lambda_a\lambda_b)=2\delta_{ab}$. Here we keep the
leading terms at large $N_c$ limit except the anomaly term which
arises from the order of $1/N_c$. Note that in order to make clear
for two independent expansions, namely $1/N_c$ expansion
characterized by $p^2/\Lambda_f^2$ in the large $N_c$ limit, and
the momentum expansion described by $p^2/\Lambda_{\chi}^2$, we
have introduced a scaling factor $\Lambda_{\chi} \simeq 1$ GeV and
redefined the low energy coupling constants $L_i$ introduced in
ref. \cite{CHPT} via $L_i = \chi_i\ f^2/4\Lambda_{\chi}^2 $ and
$H_j = \kappa_j\ f^2/4\Lambda_{\chi}^2$, so that the coupling
constants $\chi_i$ ($i=3,5,8$) and $\Lambda_{\chi}$ are constants
in the large $N_c$ limit and the whole Lagrangian is multiplied by
$f^2$ and is of order $N_c$ except the U(1) anomalous term.  This
is because when applying the large $N_c$ counting rules to all
terms in Lagrangian, it shows that $L_i = O(N_c)$ ($i\neq 7$) and
$H_j = O(N_c)$. Here $L_7$ is at order of $N_c^2$. Numerically,
one sees that $\chi_i = O(1)$ for $\Lambda_{\chi} = 1$ GeV.

  As the first step, we represent the quark currents and densities by the chiral fields
 \begin{eqnarray}
\bar{q}_{jL}\gamma^{\mu} q_{iL} & \equiv &
\frac{\delta {\cal L} }{\delta(l_{\mu}(x))_{ji}}\,= -
i\frac{f^2}{2}\{ U^{\dagger}\partial^{\mu} U  \nonumber \\
& & -  r\frac{\chi_5}{2\Lambda_{\chi}^2} \left(\partial^{\mu} U^{\dagger}{\cal M}
-{\cal M}^{\dagger}\partial^{\mu} U
+\partial^{\mu} U^{\dagger} U {\cal M}^{\dagger} U - U^{\dagger}{\cal M} U^{\dagger}
\partial^{\mu} U \right) \}_{ij}\ , \\
\bar{q}_{jR} q_{iL}
& \equiv & -\frac{\delta {\cal L} }{\delta{\cal M}_{ji}}
= -r\frac{f^2}{4}\left( U^{\dagger} + \frac{\chi_5}{\Lambda_{\chi}^2}
 \partial_{\mu} U^{\dagger} \partial^{\mu} U  U^{\dagger} +
 2r \frac{\chi_8}{\Lambda_{\chi}^2} U^{\dagger}{\cal M} U^{\dagger} +
 r \frac{\kappa_2}{\Lambda_{\chi}^2} {\cal M}^{\dagger} \right)_{ij}
\end{eqnarray}
Similarly one can obtain the right-handed currents and densities. With these definitions,
all the current $\times$ current and density $\times $ density four quark operators
can be reexpressed in terms of the chiral fields, we may call such chiral representations of
four quark operators $Q_i$ as chiral operators denoted by $Q^{\chi}_i$
correspondingly. At $p^2$ order, $Q^{\chi}_i$ can be written as the following form
\begin{eqnarray}
Q_1^{\chi} + H.c. & = & - f^4\ tr\left(\lambda_6
U^{\dagger}\partial_{\mu} U \right) tr \left(\lambda^{(1)}
U^{\dagger}\partial^{\mu} U \right) + O(1/\Lambda_{\chi}^2 ) \ ,
\nonumber \\ Q_2^{\chi} + H.c. & = & - f^4\ tr\left(\lambda_6
U^{\dagger}\partial_{\mu} U \lambda^{(1)} U^{\dagger}
\partial^{\mu} U\right) + O(1/\Lambda_{\chi}^2 )  \ ,  \nonumber
\\ Q_3^{\chi} + H.c. & = & - f^4\ tr\left(\lambda_6
U^{\dagger}\partial_{\mu} U \right) tr
\left(U^{\dagger}\partial^{\mu} U \right) + O(1/\Lambda_{\chi}^2 )
\ , \nonumber \\ Q_4^{\chi} + H.c. & = & - f^4\ tr\left(\lambda_6
\partial_{\mu} U^{\dagger} \partial^{\mu} U\right) +
O(1/\Lambda_{\chi}^2 ) \ ,  \nonumber \\ Q_5^{\chi} + H.c. & = & -
f^4\ tr\left(\lambda_6 U^{\dagger}\partial_{\mu} U \right) tr
\left(U\partial^{\mu} U^{\dagger} \right) + O(1/\Lambda_{\chi}^2 )
\ , \\ Q_6^{\chi} + H.c. & = & + f^4\
\left(\frac{r^2\chi_5}{\Lambda_{\chi}^{2}}\right)
tr\left(\lambda_6 \partial_{\mu} U^{\dagger} \partial^{\mu}
U\right) + O(1/\Lambda_{\chi}^4 )  \ ,  \nonumber \\ Q_7^{\chi} +
H.c. & = & -\frac{1}{2}Q_5^{\chi}- \frac{3}{2}f^4\
tr\left(\lambda_6 U^{\dagger}\partial_{\mu} U \right) tr
\left(\lambda^{(1)} U\partial^{\mu} U^{\dagger} \right) +
O(1/\Lambda_{\chi}^2 ) \ , \nonumber \\ Q_8^{\chi} + H.c. & = &
-\frac{1}{2} Q_6^{\chi} + f^4 r^{2}\frac{3}{4}\, tr\left(\lambda_6
U^{\dagger} \lambda^{(1)} U \right) \nonumber \\
 &+&f^4 r^{2}\frac{3}{4}\frac{\chi_5}{\Lambda_{\chi}^{2}} tr\lambda_6
 \left(U^{\dagger}\lambda^{(1)}U \partial_{\mu}U^{\dagger}\partial^{\mu} U
 +\partial_{\mu} U^{\dagger}\partial^{\mu}U
 U^{\dagger}\lambda^{(1)}U\right) \nonumber \\
 &+& f^4 r^{2}\frac{3}{4}\frac{\chi_8 }{\Lambda_{\chi}^{2}}\, 2r\,
 tr\lambda_6 \left( U^{\dagger}\lambda^{(1)} U {\cal M}^{\dagger} U
 +U^{\dagger}{\cal M}
U^{\dagger}\lambda^{(1)}U \right)
+O(1/\Lambda_{\chi}^4).\nonumber
 \end{eqnarray}
with the matrix $\lambda^{(1)}$=diag.(1,0,0). Thus loop
contributions of the chiral operators $Q^{\chi}_i$ can be
systematically calculated by using ChPT with functional cut-off
momentum.

 For $K\rightarrow \pi \pi$ decay amplitudes and direct CP-violating parameter
$\varepsilon'/\varepsilon$, the most important chiral operators
are $Q^{\chi}_1$, $Q^{\chi}_2$, $Q^{\chi}_6$ and $Q^{\chi}_8$. In
fact, the chiral operators $Q^{\chi}_3$ and $Q^{\chi}_5$ decouples
from the loop evaluations at the $p^2$ order\cite{YLW1}, i.e.
\begin{equation}
Q_5^{\chi} =  Q_3^{\chi}  = 0
\end{equation}
which can explicitly be seen from the above chiral representations
due to the traceless factor $tr \left(U\partial^{\mu} U^{\dagger}
\right) = 0$ when ignoring the singlet U(1) nonet term which is
irrelevant to the Kaon decays. Here $U\partial^{\mu} U^{\dagger} =
A_{\mu}^a \lambda^a$ may be regarded as a pure gauge.  This
feature may also be understood as the fact that $Q_3$ and $Q_5$
operators are generated from strong penguin diagrams and
suppressed by $1/N_c$ factor in comparison with the operators
$Q_4$ and $Q_6$, thus in the large $N_c$ limit, they decouple
automatically. As a consequence, it implies that at the lowest
order of $p^2$, we arrive at two additional algebraic chiral
relations
 \begin{equation}
Q_4^{\chi} = Q_2^{\chi} - Q_1^{\chi}  = -f^4\,
tr\left(\lambda_6 \partial_{\mu} U^{\dagger} \partial^{\mu} U\right)
+ O(1/\Lambda_{\chi}^2) \,.
\end{equation}
and
\begin{equation}
Q_6^{\chi}  =  - \left(\frac{r^2\chi_5}{\Lambda_{\chi}^{2}}\right)
\left( Q_2^{\chi} - Q_1^{\chi} \right) = \left(\frac{r^2\chi_5}{\Lambda_{\chi}^{2}}\right)
f^4\,
tr\left(\lambda_6 \partial_{\mu} U^{\dagger} \partial^{\mu} U\right)
\end{equation}
Notice that the mass parameter $r$ is at the same order of the
energy scale $\Lambda_{\chi}$, and $\chi_5$ is at order of unit,
thus the leading non-zero contribution of $Q_6^{\chi}$ is at the
same order of  $Q_2^{\chi}$ and $Q_1^{\chi}$.

The above algebraic chiral relations were first derived in
ref.\cite{YLW1}, they have also been checked from an explicit
calculation up to the chiral one-loop level by using the usual
cut-off regularization\cite{BBG} . If naively identifying the
cut-off momentum $M$ to the QCD running scale $\mu$, the above
algebraic chiral Wu-relations, as commented by Buras, Jamin and
Lautenbacher\cite{BJL}, seem to hold only at one point when
matching to QCD. Thus  two questions have been arised:
\begin{itemize}
\item From which energy scale and up to which order of
 chiral loop corrections the algebraic chiral Wu-relations hold;
\item How can the algebraic chiral Wu-relations survive when matching
the ultraviolet cut-off momentum of ChPT to the infrared cut-off
momentum of perturbatve QCD.
\end{itemize}
Let us briefly issue the first question and leave the second
question to next section. The answer to the first question is
manifest, the algebraic chiral Wu-relations hold starting from the
energy scale where the low energy dynamics of QCD is considered to
be described by the ChPT with functional cut-off momentum. They
even survive when we include chiral-loop corrections generated
from the lowest $p^2$ order terms as Wu-relations are the
algebraic chiral operator relations, they should not be modified
by the chiral loops of the strong interactions. Note that the
coupling constants must also be replaced by the renormalized ones
at the same order of $1/N_c$ or $1/\Lambda_F^2$. The reason is
simple as the non-trival structures of $Q_6^{\chi}$ and
$(Q_2^{\chi}-Q_1^{\chi})$ at the order of $p^2$ and $p^2/N_c$ (or
$p^2/\Lambda_F^2$) are unique. The order of $p^4$ terms are
suppressed by the factors $m_K^2/\Lambda_{\chi}^2$ and
$m_{\pi}^2/\Lambda_{\chi}^2$. Therefore, in the chiral limit,
namely $m_K^2$, $m_{\pi}^2 \ll \Lambda_{\chi}^2$, the above
algebraic chiral operator relations should hold up to the order of
$p^2/N_c$ and $p^4$. This may be understood in an analogous way to
QCD, where relations of the quark operators survive from all order
of QCD corrections. The reason is simply due to that QCD is a
renormalizable theory. For the ChPT, though it is an effective
theory and not a renormalizable one in the usual sense, but it can
be constructed to be a consistent theory order by order in the
expansion of momentum and quark mass as well as $1/N_c$. Thus, to
a given order of expansion, ChPT may be regarded as a
renormalizable one in the more general sense\cite{GW}. Therefore,
the algebraic chiral operator relations must survive, at least, up
to the one-loop corrections, which has actually been checked from
our explicit calculations. It was based on this observation, we
came to our early conclusion that the direct CP-violating
parameter $\varepsilon'/\varepsilon$ can be large enough to be
measured and its favorable numerical value is likely to be around
$\varepsilon'/\varepsilon \sim (10-30)\times 10^{-4}$
\cite{YLW1,HPSW}. As the matching to QCD was not completely
considered, our previous results\cite{YLW1,HPSW} strongly depend
on the strange quark mass.

\section{Matching Between QCD and ChPT}

 Let us begin with the short-distance operator evolution from perturbative QCD.
 When the energy scale $\mu$ is high, $m_W > \mu > m_b$, there are eleven
 independent operators $Q_i$ ($i=1,\cdots, 11$). When the energy scale $\mu$
 runs down to below the bottom quark mass $m_b$ and above the charm
 quark mass $m_c$, i.e., $m_b > \mu > m_c$, the operator $Q_{11}$ decouples and
 operator $Q_{10}$ is given by the linear combination $Q_{10} = -2Q_1 + 2Q_2
 + Q_3 - Q_4$. Once the energy scale $\mu$
 goes down to below $m_c$ but above the confining scale or the energy
 scale $\Lambda_{\chi}$, i.e., $m_c > \mu > \Lambda_{\chi}$, two
 operators $Q_9$ and $Q_4$ become no longer independent and are given by the
 linear combination $Q_9 = Q_2 + Q_1$ and $Q_4 = Q_3 + Q_2 - Q_1$. Thus there
 are only seven independent operators below $m_c$ and above $\Lambda_{\chi}$.
 In order to match to the long-distance evolution of the operators,
 let us present one-loop QCD corrections of the quark operators at
 the energy scale just above the energy scale $\Lambda_{\chi}$
\begin{eqnarray}
Q_1(\mu_Q) & = & Q_1(\mu) - 3 \frac{\alpha_s}{4\pi}\ln (\frac{\mu_Q^2}{\mu^2})\,
Q_2(\mu) + O(1/N_c)\,   , \\
Q_2(\mu_Q) & = & Q_2(\mu) - 3 \frac{\alpha_s}{4\pi}\ln (\frac{\mu_Q^2}{\mu^2})\,
Q_1(\mu) \nonumber \\
 & - & \frac{1}{3}\, \frac{\alpha_s}{4\pi}\ln (\frac{\mu_Q^2}{\mu^2})\, Q_4(\mu)
- \frac{1}{3}\, \frac{\alpha_s}{4\pi}\ln (\frac{\mu_Q^2}{\mu^2})\, Q_6(\mu) + O(1/N_c)\,  , \\
Q_4(\mu_Q) & = & Q_4(\mu) - 3 \frac{\alpha_s}{4\pi}\ln (\frac{\mu_Q^2}{\mu^2})\,
Q_3(\mu) \nonumber \\
& - & \frac{\alpha_s}{4\pi}\ln (\frac{\mu_Q^2}{\mu^2})\, Q_4(\mu)
-  \frac{\alpha_s}{4\pi}\ln (\frac{\mu_Q^2}{\mu^2})\, Q_6(\mu) + O(1/N_c)\, , \\
Q_6(\mu_Q) & = & Q_6(\mu)
- \frac{\alpha_s}{4\pi}\ln (\frac{\mu_Q^2}{\mu^2})\, Q_4(\mu)
\nonumber \\
 & + & [3(N_c -1/N_c)- 1] \frac{\alpha_s}{4\pi}\ln (\frac{\mu_Q^2}{\mu^2})\,  \, Q_6(\mu)
+ O(1/N_c)\, ,\\
Q_8(\mu_Q) & = & Q_8(\mu)  +
[3(N_c -1/N_c)- 1] \frac{\alpha_s}{4\pi}\ln (\frac{\mu_Q^2}{\mu^2})\,  \, Q_8(\mu) \  ,
\end{eqnarray}
and
\begin{eqnarray}
Q_3(\mu_Q) & = & Q_3(\mu)  - \frac{11}{3}\, \frac{\alpha_s}{4\pi}
\ln (\frac{\mu_Q^2}{\mu^2})\, Q_4(\mu) - \frac{2}{3}\, \frac{\alpha_s}{4\pi}
\ln (\frac{\mu_Q^2}{\mu^2})\, Q_6(\mu) + O(1/N_c)\, , \\
Q_5(\mu_Q) & = & Q_5(\mu)  + 3\frac{\alpha_s}{4\pi}
\ln (\frac{\mu_Q^2}{\mu^2})\, Q_6(\mu) + O(1/N_c)\ ,  \\
Q_7(\mu_Q) & = & Q_7(\mu)  + 3\frac{\alpha_s}{4\pi}
\ln (\frac{\mu_Q^2}{\mu^2})\, Q_8(\mu) + O(1/N_c)\ .
\end{eqnarray}
From the above results, we come to the following observations:

i) In the large $N_c$ limit, $Q_1$, $Q_2$, $Q_4$, and $Q_6$ form
a complete set of operators under QCD corrections.

ii) The evolution of $Q_8$ is independent of other operators and only caused
by loop corrections of the density.

iii) The operator $Q_3$ is given by the linear combination $Q_3 = Q_4 -(Q_2 -Q_1)$.
The operator $Q_5$ is driven by the operator $Q_6$, and the operator $Q_7$
is driven by the operator $Q_8$.

When the energy scale $\mu$ approaches to the confining scale,
or $\mu < \Lambda_{\chi}\sim \Lambda_F \sim 1$GeV,
as we have discussed in the above sections,
long-distance effects have to be considered. The evolution of the
operators $Q_i(\mu)$ when $\mu < \Lambda_{\chi}$ is supposed
to be carried out by the one of the chiral operators
$Q_i^{\chi}(M(\mu))$ in the framework of the functional cut-off ChPT truncated to the
pseudoscalars. To be treated at the same approximations made in the short-distance operator
evolution of QCD, we should only keep the leading terms (i.e., quadratic terms of
functional cut-off momentum) and take the chiral limit, i.e., $m_K^2, m_{\pi}^2 << \Lambda_F^2$.
In such a leading $1/N_c$ approximation and chiral limit, we find that the evolution of
the operators $Q_1^{\chi}$ and $Q_2^{\chi}$
can be simply given by the following forms when the functional cut-off momentum runs
from $M(\mu)$ down to $M(\mu')$
\begin{eqnarray}
Q_1(\mu)& \rightarrow & Q_1^{\chi}(M(\mu))= Q_1^{\chi}(M(\mu'))
-\frac{2(M^2(\mu)-M^2(\mu'))}{\Lambda_F^2}
\, Q_2^{\chi}(M(\mu'))\, ,\\
Q_2(\mu)& \rightarrow & Q_2^{\chi}(M(\mu))= Q_2^{\chi}(M(\mu'))
-\frac{2(M^2(\mu)-M^2(\mu'))}{\Lambda_F^2}
\, Q_1^{\chi}(M(\mu')) \nonumber \\
& + & \frac{M^2(\mu)-M^2(\mu')}{\Lambda_F^2}\, ( Q_2^{\chi} - Q_1^{\chi})(M(\mu'))\, ,
\end{eqnarray}
where $\Lambda_F = 4\pi F =1.16$ GeV with $F$ the renormalized one of $f$.
Notice that the operators $Q_3^{\chi}$ and $Q_5^{\chi}$ decouple from the evolution, namely
$Q_3^{\chi}=0$ and $Q_5^{\chi}=0$. The results for the operators $Q_i^{\chi}$ ($i=4,6,8$) can
be written as follows
\begin{eqnarray}
Q_4(\mu)& \rightarrow & Q_4^{\chi}(M(\mu)) = (Q_2^{\chi} - Q_1^{\chi})(M(\mu))  \,  ,\\
Q_6(\mu)& \rightarrow & Q_6^{\chi}(\mu, M(\mu))= \left(1 +
 3(N_c -1/N_c)\frac{\alpha_s}{4\pi}\ln (\frac{\mu^{2}}{\mu_{\chi}^{2}}) \right)\,
Q_6^{\chi}(\mu_{\chi}, M(\mu))\, , \\
Q_8(\mu)& \rightarrow & Q_8^{\chi}(\mu, M(\mu))= \left(1 +
 3(N_c -1/N_c)\frac{\alpha_s}{4\pi}\ln (\frac{\mu^{2}}{\mu_{\chi}^{2}}) \right)\,
Q_8^{\chi}(\mu_{\chi}, M(\mu))\, .
\end{eqnarray}
where the explicit $\mu$-dependence of the operators
$Q_6^{\chi}(\mu, M(\mu))$ and $Q_8^{\chi}(\mu, M(\mu))$ arise from
the running quark mass and behavors like $1/(m_s(\mu) +
\hat{m}(\mu))^{2}$. Notice that the independent operators are
reduced once more in the long-distance operator evolution when
$\mu < \Lambda_{\chi}$ due to the algebraic chiral operator
relations. Let us now compare and match the loop results evaluated
from QCD with the ones from the ChPT with functional cut-off
momentum at the energy scale $\Lambda_{\chi}$. Substituting
eqs.(20) and (21) into eq.(12), keeping the leading $1/N_c$ terms,
we obtain, from the requirement of $\mu$-independence in the large
$N_c$ limit, i.e., $\frac{\partial}{\partial \mu} Q_1(\mu_Q)=0$,
the first matching condition,
\begin{equation}
\mu \frac{\partial}{\partial \mu} \left( \frac{2M^2(\mu)}{\Lambda_F^2}\right)
= \frac{3\alpha_s}{2\pi}\, ,
\end{equation}
which can in general be obtained by requiring the matching
between the anomalous dimensions of quark operators $Q_i(\mu)$ in QCD and the ones of
the corresponding chiral operators $ Q_i^{\chi}(M(\mu))$ in ChPT in the large $N_c$
limit, i.e.,
\begin{equation}
\gamma_i^{Meson} \equiv  \mu \frac{\partial}{\partial \mu} Q_i^{\chi}(M(\mu)) =
\gamma_i^{Quark} \equiv  \mu \frac{\partial}{\partial \mu} Q_i(\mu)\, .
\end{equation}

Analogously, substituting eqs.(20)-(23) into eq.(13), keeping the leading $1/N_c$ terms and
adopting the above first matching condition, we arrive at the second matching condition
\begin{equation}
Q_6^{\chi}(\mu_{\chi}, M(\mu)) =  -\frac{11}{2}(Q_2^{\chi} - Q_1^{\chi} ) (M(\mu)) \, ,
\quad \mu < \Lambda_{\chi}
\end{equation}
Note that such a matching condition holds for whole energy scale $\mu < \Lambda_{\chi}$. At the
special point $M(\mu)=0$, it covers the condition first presented in ref.\cite{FG}. In fact, the
above two matching conditions may simply seen by comparing eqs.(20) and (21) with eqs.(12) and (13).
On the other hand, from the chiral representation of operators and their chiral loop corrections,
we have the following chiral relation in the leading $1/N_c$ approximation and chiral limit
\begin{eqnarray}
& & Q_6^{\chi}(\mu_{\chi}, M(\mu)) \simeq
\left(- \frac{R^2_{\chi} \chi_5 ^r}{\Lambda_{\chi}^2}\right)
( Q_2^{\chi} - Q_1^{\chi})(M(\mu))\, , \quad \mu < \Lambda_{\chi} \\
& & R_{\chi} \equiv R(\mu\simeq \mu_{\chi} )\simeq m_{\pi}^{2}/\hat{m}(\mu_{\chi})
\simeq 2m_{K}^{2}/(m_s + \hat{m})(\mu_{\chi})\,  ,
\end{eqnarray}
where we have simply replaced the coupling constants $\chi_5$ and $r$ by the corresponding
renormalized ones $\chi_5 ^r$ and $R(\mu)$ as their loop corrections are
at the subleading order. When combining the second matching condition, it
allows us to fix the strange quark mass
\begin{equation}
\frac{R^2_{\chi} \chi_5 ^r}{\Lambda_{\chi}^2}  = \frac{11}{2}\,  \quad \rightarrow \quad
m_s(\mu_{\chi}) \simeq 196 MeV\, ,
\end{equation}
Here we have used the result $\Lambda_{\chi} = 1.03 \sqrt{\chi_5^r}$ GeV which is fixed
from the ratio of the kaon and pion decay constants.

The first matching condition can be rewritten as follows after integration
\begin{equation}
\frac{2M^2(\mu)}{\Lambda_F^2} = \frac{2M^2_0}{\Lambda_F^2} + \frac{3\alpha_s}{4\pi}\,
\ln(\frac{\mu^2}{\mu^2_0} )\, ,
\end{equation}
where $\mu_0$ and $M_0 \equiv M(\mu = \mu_0)$ are two integral
constants. It is seen that the $\mu$-dependence of the functional
cut-off momentum $M(\mu)$ is now logarithmic. Noticing the
approximation $\ln (\mu^2/\mu_0^2) \simeq \mu^2/\mu^2_0 - 1$ when
$\mu^2 \sim \mu^2_0$, namely the functional cut-off momentum
$M(\mu)$ is approximately proportional to $\mu$ when $\mu$ runs
down and approaches to the low energy scale $\mu_0$ which is
expected to be slightly above the QCD scale $\Lambda_{QCD}$, we
then have $M^2_0 = (3\alpha_s(\mu_0)/8\pi)\Lambda_F^2 $. Thus the
$\mu$-dependence of the functional cut-off momentum $M(\mu)$ can
be written as
\begin{equation}
\frac{2M^2(\mu)}{\Lambda_F^2} \simeq \frac{3\alpha_s}{4\pi}
 + \frac{3\alpha_s}{4\pi}\, \ln(\frac{\mu^2}{\mu_0^2} )\, ,
 \end{equation}
which shows that after imposing the matching condition for the
anomalous dimensions between quark operators $Q_i(\mu)$ in QCD and
the corresponding chiral operators $ Q_i^{\chi}(M(\mu))$ in the
ChPT with functional cut-off momentum, the dimensionless ratio
$M^2/\Lambda_F^2$ is only related to the strong coupling constant
$\alpha_s$ and becomes scheme-independent, which implies that the
long-distance operator evolution in  ChPT with functional cut-off
momentum can be carried out by using any approach. For instance,
with and without separating factorized and non-factorized
contributions, should obtain the same results after appropriately
considering the matching between QCD and ChPT with functional
cut-off momentum.

In general, we have $\mu_0 > \Lambda_{QCD}$. To fix the value of
$\mu_0$, we use $M_0 \simeq \mu_0$. Thus $\mu_0$ (or
$\alpha_s(\mu_0)$ ) is determined via
\begin{equation}
\mu_0 \simeq \Lambda_F \sqrt{3\alpha_s(\mu_0)/8\pi}
\end{equation}
 Using the definition $\alpha_s(\mu) = 6\pi /[(33-2n_f)
 \ln (\mu^2/\Lambda_{QCD}^2)]$ with $n_f =3$,
the initial low energy scale $\mu_0$ is found, for $\Lambda_{QCD}
= 325\pm 80$ MeV, to be
\begin{equation}
 \mu_0 \simeq 435\pm 70 MeV
  \quad or \quad  \alpha_s(\mu_0)/2\pi \simeq 0.19^{+0.06}_{-0.05}     \  .
\end{equation}
With such an initial value of $\mu_0$, the functional cut-off
momentum $M(\mu)$ at $\mu = \Lambda_{\chi}$ yields the following
corresponding value
\begin{equation}
M_{\chi} \equiv M(\mu = \Lambda_{\chi} \simeq 1 GeV) \simeq
0.71^{+0.11}_{-0.12} GeV \ .
\end{equation}
which provides the possible allowed range of the energy scale
where the ChPT with functional cut-off momentum can be used to
describe the low energy behavior of QCD at large $N_c$ limit.

\section{Evolution of Long-distance Chiral Operators}

 From the above analyzes, the $\Delta S = 1$ low energy ($\mu < \Lambda_{\chi}$)
 effective Hamiltonian for calculating $K\rightarrow \pi \pi$ decay amplitudes may be
written as
\begin{equation}
{\cal H}_{eff}^{ \Delta S=1}= \frac{G_F}{\sqrt{2}}
\;\lambda_u\sum_{i=1,2,4,6,8} c_i(\mu)\,Q_i^{\chi}(M(\mu))\ ,
 \quad (\mu < \Lambda_{\chi} )
\end{equation}
 we may now adopt the matching conditions and algebraic chiral operator relations
  to investigate the evolution of the chiral operators $Q_i^{\chi}(M(\Lambda_{\chi}))$.
 The first matching condition enables us to sum over all the leading terms via
 renormalization group equation down to the energy scale $\mu_0$, and the second
  matching condition together with
 the algebraic chiral operator relations allows us to evaluate the penguin operators
$Q_4^{\chi}(M)$ and $Q_6^{\chi}(M)$ from the operators $Q_1^{\chi}(M)$ and $Q_2^{\chi}(M)$.
So that the operators $Q_1^{\chi}(M)$ and $Q_2^{\chi}(M)$ form a complete set for the operator
evolution below the energy scale $\mu \simeq \Lambda_{\chi} \simeq 1 $ GeV, or correspondingly,
 below the functional cut-off momentum $M(\mu\simeq \Lambda_{\chi})
  \simeq 0.71^{+0.11}_{-0.12} $ GeV for $\Lambda_{QCD} = 325 \pm 80$ MeV.
 It is convenient to choose a new operator basis
 $Q_{\pm}^{\chi}(M(\mu)) = Q_2^{\chi}(M(\mu))\pm Q_2^{\chi}(M(\mu))$.
 The anomalous dimension matrix for the
 basis $(Q_{-}^{\chi}, Q_{+}^{\chi})$ is found to be
 \begin{equation}
 \gamma = \frac{\alpha_{s}}{2\pi} \left( \begin{array}{cc}
 -9/2 & 0 \\
 -3/2 & 3 \\
 \end{array} \right)
 \end{equation}
  Following the standard procedure of the renormalization group evolution
   with the initial conditions for the Wilson coefficient functions:
  $c_{-}(\Lambda_{\chi}) = c_2 (\Lambda_{\chi}) - c_1 (\Lambda_{\chi})$ and
  $c_{+}(\Lambda_{\chi}) = c_2 (\Lambda_{\chi}) + c_1(\Lambda_{\chi}) $ ,
  we find in the leading logarithmic approximation that
  \begin{eqnarray}
  Q_{-}^{\chi}(M(\Lambda_{\chi})) & = & \eta_{\chi} ^{-1/2}\
   Q_{-}^{\chi}(\mu_0)\ , \\
  Q_{+}^{\chi}(M(\Lambda_{\chi})) & = &
  \eta_{\chi}^{1/3}Q_{+}^{\chi}(\mu_0) + \frac{1}{5}\left(
  \eta_{\chi}^{-1/2} - \eta_{\chi}^{1/3} \right)
  Q_{-}^{\chi}(\mu_0)\  ,
  \end{eqnarray}
  with $\eta_{\chi} = \alpha_{s}(\Lambda_{\chi})/\alpha_{s}(\mu_0)$, and
  \begin{eqnarray}
  Q_{-}^{\chi}(\mu_0) & = & Q_{-}^{\chi}(0) + \frac{9\alpha_s(\mu_0)}{8\pi}\
  Q_{-}^{\chi}(0)\  , \\
  Q_{+}^{\chi}(\mu_0) & = & Q_{+}^{\chi}(0) -  \frac{3\alpha_s(\mu_0)}{4\pi}\
  Q_{+}^{\chi}(0) +  \frac{3\alpha_s(\mu_0)}{8\pi}\  Q_{-}^{\chi}(0)\  ,
  \end{eqnarray}

   In the above analyzes, we have taken the chiral limit $m_K^2,
   m_{\pi}^2 \ll\Lambda_F^2$. From the chiral one loop results,
   the finite meson mass contributions can be approximately included
   by modifying the above results into the following form
  \begin{eqnarray}
  Q_{-}^{\chi}(M(\Lambda_{\chi})) & = & \eta_{\chi} ^{-1/2}\,
  \eta_{-}(M_{\chi})\, Q_{-}^{\chi}(\mu_0)\ , \\
  Q_{+}^{\chi}(M(\Lambda_{\chi})) & = &
  \eta_{\chi}^{1/3}\, \eta_1(M_{\chi})\, Q_{+}^{\chi}(\mu_0) + \frac{1}{5}\left(
  \eta_{\chi}^{-1/2} - \eta_{\chi}^{1/3} \right)\, \eta_2(M_{\chi})\, Q_{-}^{\chi}(\mu_0)\  ,
  \end{eqnarray}
  with
\begin{eqnarray}
\eta_{-}(M_{\chi})& \simeq & 1 + \frac{\frac{3}{4}m_K^2 -
\frac{9}{2}m_{\pi}^2}{\Lambda_{F}^2}
 \ln \left(1 + \frac{M^2(\mu)}{\tilde{m}^2}\right) \, , \nonumber \\
\eta_{1}(M_{\chi})& \simeq & 1 + \frac{\frac{1}{4}m_K^2 +
3m_{\pi}^2}{\Lambda_{F}^2}
 \ln \left(1 + \frac{M^2(\mu)}{\tilde{m}^2}\right) \, , \nonumber \\
 \eta_{2}(M_{\chi})& \simeq & 1 + \frac{m_K^2 -
\frac{3}{2}m_{\pi}^2}{M_{\chi}^2}
 \ln \left(1 + \frac{M^2(\mu)}{\tilde{m}^2}\right) \, .
\end{eqnarray}
Numerically, we use $\tilde{m}\simeq 300$MeV, $m_{K} = 0.495$ GeV
and $m_{\pi} = 0.137$GeV. When the QCD scale takes the value
$\Lambda_{QCD} = 325\pm 80$ MeV with the corresponding
  low energy cut-off momentum $\mu_0 \simeq 435\pm 70$ MeV, we have
  \begin{eqnarray}
  Q_{-}^{\chi}(M(\Lambda_{\chi})) & = & (3.17^{+0.66}_{-0.43})\  Q_{-}^{\chi}(0)
  =  Q_{4}^{\chi}(M(\Lambda_{\chi}))\  ,  \\
  Q_{+}^{\chi}(M(\Lambda_{\chi})) & = & (0.55^{-0.09}_{+0.06})\ Q_{+}^{\chi}(0)
  + (0.8^{+0.11}_{-0.05})\ Q_{-}^{\chi}(0)\  , \\
  Q_{6}^{\chi}(\mu_{\chi}, M(\Lambda_{\chi})) & = & -\frac{11}{2}
   Q_{-}^{\chi}(M(\Lambda_{\chi}))
   = - (17.44^{+3.62}_{-2.37})\  Q_{-}^{\chi}(0)\  , \\
  Q_{8}^{\chi}(\mu_{\chi}, M(\Lambda_{\chi})) & = & \frac{33}{8}\
  \frac{\Lambda_{\chi}^{2}}{\chi_5^{r}(m_K^2 - m_{\pi}^2 )}\
  ( Q_{+}^{\chi} + Q_{-}^{\chi} )(0) =
  19.18\ ( Q_{+}^{\chi} + Q_{-}^{\chi} )(0)\ .
  \end{eqnarray}
   Which show that the isospin $I = 2$ amplitude $A_2$ is suppressed by a factor of about 2
   as it only receives contributions
   from the operator $Q_{+}^{\chi}(M(\Lambda_{\chi}))$, while the isospin $I =0 $ amplitude $A_0$
   is enhanced by a large factor as it mainly gets contributions from the operator
   $Q_{-}^{\chi}(M(\Lambda_{\chi}))$. On the other hand,
   the direct CP-violating parameter $\varepsilon'/\varepsilon$
  is expected to be large since the significant enhancement
   of $Q_{6}^{\chi}(\mu_{\chi}, M(\Lambda_{\chi}))$ relative to
   $Q_{8}^{\chi}(\mu_{\chi}, M(\Lambda_{\chi}))$
   is seen to be resulted from the algebraic chiral operator relation and matching condition.
   We are going to present our numerical predictions for the isospin amplitudes
   and the direct CP-violating parameter
    $\varepsilon'/\varepsilon$ in the next section.

    \section{ Predictions for $\varepsilon'/\varepsilon$  and $\Delta I = 1/2$ Rule}

 We are now in the position to calculate the $K\rightarrow \pi \pi$ decay
 amplitudes $A_I$ with isospin $I$
 \begin{equation}
 A_I \cos\delta_{I} = \langle \pi\pi | {\cal H}_{ef\ f}^{ \Delta S=1} | K \rangle \equiv
 \frac{G_F}{\sqrt{2}} \;\lambda_u\sum_{i=1,2,4,6,8} c_i(\Lambda_{\chi})
 Re\langle Q_i^{\chi}(M(\Lambda_{\chi}))\rangle_I
 \end{equation}
 The CP-conserving amplitudes are given by
 \begin{eqnarray}
 Re A_0 \cos\delta_{0}& = & \frac{G_F}{\sqrt{2}} \; Re\lambda_u\sum_{i=1,2,4,6,8}
 z_i(\Lambda_{\chi})Re\langle Q_i^{\chi}(M(\Lambda_{\chi}))\rangle_0 \nonumber \\
 & \simeq & \frac{G_F}{\sqrt{2}} \; Re\lambda_u [
 \frac{1}{2}z_{-}(\Lambda_{\chi})Re\langle Q_{-}^{\chi}(M(\Lambda_{\chi}))\rangle_0 +
 \frac{1}{2}z_{+}(\Lambda_{\chi})Re\langle Q_{+}^{\chi}(M(\Lambda_{\chi}))\rangle_0 \nonumber \\
 & + & z_{4}(\Lambda_{\chi})Re\langle Q_{4}^{\chi}(M(\Lambda_{\chi}))\rangle_0 +
 z_{6}(\Lambda_{\chi})Re\langle Q_{6}^{\chi}(M(\Lambda_{\chi}))\rangle_0 ] \  , \\
 Re A_2 \cos\delta_{2} & = & \frac{G_F}{\sqrt{2}} \; Re\lambda_u\sum_{i=1,2,8}
 z_i(\Lambda_{\chi})Re\langle Q_i^{\chi}(M(\Lambda_{\chi}))\rangle_2 \\
 & \simeq & \frac{G_F}{\sqrt{2}} \; Re\lambda_u [
 \frac{1}{2}z_{-}(\Lambda_{\chi})Re\langle Q_{-}^{\chi}(M(\Lambda_{\chi}))\rangle_2 +
 \frac{1}{2}z_{+}(\Lambda_{\chi})Re\langle Q_{+}^{\chi}(M(\Lambda_{\chi}))\rangle_2 ]
  \  , \nonumber
 \end{eqnarray}
 and the CP-violating amplitudes are dominated by
 $\langle Q_6^{\chi}(M(\Lambda_{\chi})\rangle_0$
 and $\langle Q_8^{\chi}(M(\Lambda_{\chi})\rangle_2$
 \begin{eqnarray}
 Im A_0 \cos\delta_{0} & = & -\frac{G_F}{\sqrt{2}} \; Im\lambda_t\sum_{i=1,2,4,6,8}
 y_i(\Lambda_{\chi})Re\langle Q_i^{\chi}(M(\Lambda_{\chi}))\rangle_0 \nonumber \\
 & \simeq & -\frac{G_F}{\sqrt{2}} \; Im\lambda_t \  [
 y_{6}(\Lambda_{\chi})Re\langle Q_{6}^{\chi}(M(\Lambda_{\chi}))\rangle_0 ]\  , \\
 Im A_2 \cos\delta_{2} & = & -\frac{G_F}{\sqrt{2}} \; Im\lambda_t \sum_{i=1,2,8}
 y_i(\Lambda_{\chi})Re\langle Q_i^{\chi}(M(\Lambda_{\chi}))\rangle_2 \nonumber \\
 & \simeq & \frac{G_F}{\sqrt{2}} \; Im \lambda_t \  [
 y_{8}(\Lambda_{\chi})Re\langle Q_{8}^{\chi}(M(\Lambda_{\chi}))\rangle_2 ] \  .
 \end{eqnarray}
 From the definition of direct CP-violating parameter $\varepsilon'/\varepsilon$
  \begin{equation}
  \frac{\varepsilon'}{\varepsilon} = \frac{\omega}{\sqrt{2}|\varepsilon|}
  \left( \frac{Im A_2}{Re A_2} -
  \frac{Im A_0}{Re A_0} \right)
  \end{equation}
 with $\omega = Re A_2/ Re A_0 = 1/22.2$, we arrive at the following general expression
 \begin{equation}
  \frac{\varepsilon'}{\varepsilon} = \frac{G_F}{2}\frac{\omega}{|\varepsilon|Re A_0 }
  Im \lambda_t \left( h_0 - h_2/\omega \right)
  \end{equation}
  Here $h_0$ and $h_2$ are given by the isospin $I=0$ and $I=2 $ hadronic matrix elements of
 relevant operators
  \begin{eqnarray}
  h_0 & = & (\cos\delta_0)^{-1}\sum_{i=1,2,4,6,8} y_i(\Lambda_{\chi})
  Re\langle Q_i^{\chi}\left(M(\Lambda_{\chi})\right)\rangle_0
  \left(1 - \Omega_{IB} \right) \nonumber \\
  & \simeq & (\cos\delta_0)^{-1} y_6(\Lambda_{\chi})
  Re\langle Q_6^{\chi}\left(M(\Lambda_{\chi})\right)\rangle_0
   \left(1 - \Omega_{\eta + \eta'} \right)\  , \\
  h_2 & = & (\cos\delta_2)^{-1}\sum_{i=1,2,8} y_i(\Lambda_{\chi})
  Re\langle Q_i^{\chi}\left(M(\Lambda_{\chi})\right)\rangle_2  \nonumber \\
  &\simeq & (\cos\delta_2)^{-1} y_8(\Lambda_{\chi})
  Re\langle Q_8^{\chi}\left(M(\Lambda_{\chi})\right)\rangle_2 \  ,
  \end{eqnarray}
 where we have taken into account the possible isospin breaking effect
$\Omega_{IB}$, its previously estimated value was $\Omega_{IB}
\simeq 0.25 \pm 0.1$\cite{ISB}. The most recent refined
calculation in\cite{ISB1} gives a smaller value $\Omega_{IB}
\simeq 0.16 \pm 0.03$ but with a large error\cite{ISB2}. In our
present numerical calculations, we use $\Omega_{IB} \simeq 0.16$.
The CKM factors $Re\lambda_u$ and $Im\lambda_t$ are given in the
Wolfenstein parameterization\cite{LW} as follows
 \begin{equation}
 Re\lambda_u = Re(V_{us}^{\ast}V_{ud})= \lambda\ , \quad
Im\lambda_t = Im(V_{ts}^{\ast}V_{ta}) = A^2\lambda^5 \eta
 \end{equation}

  To evaluate the numerical results, we are going to take the following reliable values for all
  relevant parameters. For the involved energy scales, we have: $\Lambda_{QCD} = 325\pm 80$ MeV,
  $\mu_{0} = 435\pm 70 $MeV, $\Lambda_{\chi}=1.0$GeV and $\Lambda_F = 1.16$ GeV. For the
  Wilson coefficient functions, we only use the leading order results at one-loop level
  for a consistent analysis since the chiral operators have only been evaluated up to
  the leading order at the chiral one-loop level, namely at the order of $1/N_c\sim
  M^2/\Lambda_{F}^2\sim \alpha_s$ in the large $N_c$ approach.
  Their values can be read following refs. \cite{BBH,PW}.
  The numerical values at $\mu = \Lambda_{\chi}$ are regarded as the `initial conditions'
  for the chiral operator evolution and read for $\Lambda_{QCD} = 325\pm 80$ MeV :
  $z_{-}(\Lambda_{\chi}) =(z_2 - z_1)(\Lambda_{\chi}) = 2.181^{+0.197}_{-0.177}$,
  $z_{+}(\Lambda_{\chi}) = (z_2 + z_1)(\Lambda_{\chi}) = 0.685\mp 0.029$,
  $z_{4}(\Lambda_{\chi}) = -(0.012\pm 0.003)$ and $z_{6}(\Lambda_{\chi}) = -(0.013\pm 0.003)$,
  as well as $y_{6}(\Lambda_{\chi}) = -\left(0.113^{+0.024}_{-0.021}\right)$
  and $y_{8}(\Lambda_{\chi})/\alpha = 0.158^{+0.040}_{-0.033}$. The hadronic matrix
  elements of chiral operators at cut-off momentum $M=0$ take their values at the tree-level:
  $\langle Q_{-}^{\chi}(0)\rangle_0 = 36.9\times 10^6 $ MeV$^{3}$,
  $\langle Q_{+}^{\chi}(0)\rangle_0 = 12.3\times 10^6 $ MeV$^{3}$,
  $\langle Q_{+}^{\chi}(0)\rangle_2 = 34.8\times 10^6 $ MeV$^{3}$ and
  $\langle Q_{8}^{\chi}(\Lambda_{\chi}, 0)\rangle_2= 328.8\times 10^6 $ MeV$^{3}$.
  For the CKM matrix elements, there remain big uncertainties arising from the
  single CP-violating phase, two matrix elements $V_{ub}$ and $V_{cb}$,
  or the corresponding Wolfenstein parameters $\eta$, $\rho$ and $A$. For a numerical estimate,
  we take $Re\lambda_u = 0.22$ and $Im\lambda_t =1.2 \times 10^{-4}$ as the central values.
  With these input values, we obtain, in the leading $1/N_c$ approximation, the isospin amplitudes
  \begin{eqnarray}
  Re A_0 & = &  (2.56^{+0.78}_{-0.37})\, \times 10^{-4}\ (\cos\delta_0)^{-1}\ MeV
  = (3.10^{+0.94}_{-0.61})\, \times 10^{-4}\ MeV \  , \\
  Re A_2 & = &  (0.12\mp 0.02) \times 10^{-4}\ (\cos\delta_2)^{-1}\ MeV
   = (0.12\mp 0.02)\times 10^{-4}\ MeV \  ,
  \end{eqnarray}
 which agree well with the experimental data: $Re A_0 = 3.33 \times 10^{-4}$ MeV and
 $Re A_2 = 0.15 \times 10^{-4}$ MeV . Here we have used the final state interaction phases,
 $\delta_0 = (34.2\pm 2.2)^o$ and $\delta_2 = (-6.9 \pm 0.2)^o$ \cite{CO}.
 Consistently, the direct CP-violating parameter $\varepsilon'/\varepsilon$ is found,
  in the leading $1/N_c$ approximation with $\Lambda_{QCD}=325\pm 80$ MeV, to be
  \begin{equation}
  \frac{\varepsilon'}{\varepsilon} = (23.6^{+12.4}_{-7.8})\, \times 10^{-4}
  \end{equation}
  for the central value of $Im \lambda_t = 1.2 \times 10^{-4}$ resulted from fitting the indirect
  CP-violating parameter $\varepsilon$, $|V_{ub}|$, $B^0-\bar{B}^0$ and $B^0_s-\bar{B}^0_s$
  mixings. It is of interest to note that this central value also agrees with the one
 predicted from ten useful relations among the quark masses and mixing angles obtained in the
  SUSY GUT model\cite{CW}. When considering the possible allowed range for the CKM
  matrix elements extracted from fitting the present experimental data, we have
  \begin{equation}
  (13.8^{+7.2}_{-4.5})\, \times 10^{-4} \leq \frac{\varepsilon'}{\varepsilon} =
    (23.6^{+12.4}_{-7.8})\,  \times 10^{-4} \left(\frac{Im \lambda_t}{1.2 \times 10^{-4}} \right)
   \leq   (33.6^{+17.6}_{-11.1})\,  \times 10^{-4}
  \end{equation}
  for  the possible allowed range $0.7\times 10^{-4} < Im\lambda_t < 1.7 \times 10^{-4}$.
  It is noticed that the present new predictions for the isospin amplitudes and direct
  CP-violating parameter $\varepsilon'/\varepsilon$ further confirm our early
  conclusions\cite{YLW1,HPSW}. Our new predictions are consistent
  with the world average\cite{NA48N}
  \begin{equation}
  Re(\varepsilon'/\varepsilon)  =  (19.2\pm 2.4)\times 10^{-4}
  \quad (World\, \ Average\, \ 2000)
  \end{equation}
  which is obtained by taking into account the results from four
  collaboration groups. They contain two published results reported early
  by NA31 collaboration and E731 collaboration
  \begin{eqnarray}
  Re(\varepsilon'/\varepsilon) & = & (23\pm 7)\times
   10^{-4}, \quad (1993\, NA31)\cite{NA31}\, ; \nonumber \\
   Re(\varepsilon'/\varepsilon) & = & (7.4\pm 5.9)\times 10^{-4}
  \quad (1993\, E731) \cite{E731}\,
  \end{eqnarray}
 and the recent new results reported by
 the KTeV collaboration at Fermilab and the NA48 collaboration at CERN:
  \begin{eqnarray}
   Re(\varepsilon'/\varepsilon) & = & (28.0\pm 3.0\pm 2.8)\times 10^{-4}
  \quad (1999\, KTeV) \cite{KTEV}\, ; \nonumber \\
   Re(\varepsilon'/\varepsilon) & = & (18.5\pm 4.5\pm 5.8)\times
   10^{-4}, \quad (1999\, NA48)\cite{NA48}\, ; \nonumber \\
  Re(\varepsilon'/\varepsilon) & = & (14.4\pm 4.3)\times
  10^{-4} \quad (2000\, NA48)\cite{NA48N}\ ,
  \end{eqnarray}

 Before drawing our conclusions, we would like to address the following points:

 1. The main uncertainties for the predictions arise from the QCD scale $\Lambda_{QCD}$
 (or the low energy scale $\mu_0$) and the combined CKM factor
 $Im\lambda_t$. Nevertheless, the uncertainties from the energy scale may be reduced from
 comparing the predicted isospin amplitudes $A_0$ and $A_2$ with the well measured
 ones. As a consequence, it is seen from eqs.(59) and (60) that
 the results corresponding to the large values of $\Lambda_{QCD} > 325$ MeV appear not
 so favorable.

 2. From the above point of view, it is seen that from the isospin amplitude $A_2$,
 the ratio $\varepsilon'/\varepsilon$ favors the low values
\begin{equation}
\frac{\varepsilon'}{\varepsilon} = (16\pm 7)\,  \times 10^{-4}
 \end{equation}
 while from the isospin amplitude $A_0$, it favors the high values
\begin{equation}
\frac{\varepsilon'}{\varepsilon} = (24 \pm 10)\,  \times 10^{-4}
 \end{equation}
 From the ratio of the two amplitudes $ReA_0/ReA_2$ , i.e., the $\Delta I =1/2$
 rule, the ratio $\varepsilon'/\varepsilon$ favors the middle values
\begin{equation}
\frac{\varepsilon'}{\varepsilon} = (20\pm 9)\,  \times 10^{-4}
 \end{equation}
 which is consistent with the most recent results reported by the
 NA48 collaboration at CERN\cite{NA48N,NA48} and the KTeV collaboration at
 Fermilab\cite{KTEV}. In fact, it is very close to the average
 value from NA48 and KTeV. While the central values from two
 experimental groups differ from each other at 3.5-$\sigma$ level.

 3. The above results are renormalization scheme independent as the consistent matching
 between QCD and ChPT considered in the present paper is at the leading one-loop
 order of $1/N_c\sim \alpha_s \sim 1/\Lambda_F^2$
 around the scale $\Lambda_{\chi}$. The renormalization scheme
 dependence arises from the next-to-leading order of perturbative QCD, which
 could become substantial for some of the Wilson coefficient functions when the
 renormalization scale $\mu$ runs down to around the scale $\Lambda_{\chi} =
 1$GeV. In our present approach, the scheme for the long-distance evolution
 is fixed by the ChPT with functional cut-off momentum. For
 matching to this scheme, it is useful to introduce a scheme independent basis for the
 perturbative QCD calculation of short-distance physics. Then applying our above procedure to
 find out the matching conditions at the next-to-leading order
 $1/N_c^2\sim \alpha_s^2 \sim 1/\Lambda_F^4$. To work out the scheme independent basis,
 it may be helpful to adopt the method discussed in ref.\cite{WB} and use the cut-off
 momentum basis. The study of scheme independent basis in perturbative QCD is
 beyond the purposes of the present paper. Some effort is
 being made\cite{SCHEME} though it is not yet full understood.

 \section{Conclusions}

 We have simultaneously analyzed the direct CP-violating parameter
 $\varepsilon'/\varepsilon$ and $\Delta I = 1/2$ rule in kaon decays
 by considering a consistent matching scheme
 between QCD and ChPT. Our main points may be summarized as
 follows: i) Starting from the chiral Lagrangian obtained in terms
 of the momentum and quark mass expansion with low energy coupling
 constants given by the leading terms of the $1/N_c$ expansion. The
 $N_c$ behavior in the concerned chiral Lagrangian has explicitly been
 characterized by the scale factor $\Lambda_f^2 \sim N_c$; ii) The four
 quark operators for weak kaon decays at low energies have been assumed to be
 represented by the chiral operators in the large $N_c$ limit. It has also
 been shown that there is simplification in ChPT which leads to useful algebraic
 chiral operator relations. Those relations survive even when including loop
 corrections; iii) We have adopted the usual cut-off regularization scheme\cite{BBG}
 for all the diagrams with a single cut-off momentum for a systematical analysis,
 and given up the scheme of separating factorized and non-factorized parts
 with two cut-off scales; iv) The cut-off momentum $M$ has been considered to be
 the function of the QCD running scale $\mu$, i.e., $M\equiv M(\mu)$,  instead of
 naively identifying it to the perturbative QCD running scale $\mu$. The form
 of the functional cut-off momentum $M(\mu)$ has been determined via the matching
 requirement, so that the chiral loop results become scheme independent. As a
 consequence, two useful matching conditions have been obtained,
 which has allowed us to evaluate the long-distance chiral operators
 and sum over the leading non-perturbative contributions. Of particular,
 the $\Delta I = 1/2$ rule can consistently  be understood and the resulting
 direct CP-violating parameter $\varepsilon'/\varepsilon$ become large enough to be
 measured, which also confirms our early conclusions\cite{YLW1,HPSW}. Taking into account
 the simultaneous consistent analysis for the isospin amplitudes $A_0$ and $A_2$, the numerical
 result for the ratio $\varepsilon'/\varepsilon$ is in favor of
 the values
\begin{equation}
\frac{\varepsilon'}{\varepsilon} = (20\pm 9)\,  \times 10^{-4}
 \end{equation}
 which may be regarded as the favorable prediction in our present analyzes.
 The prediction is also consistent, within the theoretical and experimental uncertainties,
 with the present data\cite{NA48N,NA31,E731,KTEV,NA48}.
 Finally, we would like to remark that we have neglected in our present analyzes the
 subleading contributions, their effects are in general small and
 will be investigated elsewhere in detail.

{\bf Acknowledgements:} The author would like to express gratitude
to E.A. Paschos for his enlightening discussions about the
matching problem and for the hospitality at Dortmund University.
He would also like to thank W. Bardeen and L. Wolfenstein for
their useful discussions. This work was supported in part by the
NSF of China under Grant No. 19625514 and Chinese Academy of
Sciences. The partial support from the Bundesministerium f\"{u}r
Bildung, Wissenschaft, Forschung und Technologie (BMBF),
057D093P(7), Bonn, FRG, and DFG Antrag PA-10-1 is also
acknowledged.

 \end{document}